\documentclass[twocolumn,showpacs,preprintnumbers,prd,amsmath,amssymb]{revtex4}
\usepackage[dvips]{graphicx}
\usepackage{dcolumn}
\usepackage{bm}
\newcommand{\p}{\partial}
\newcommand{\eu}{{\rm e}}
\newcommand{\dd}{{\rm d}}

\begin{document}

\preprint{BICOCCA-FT-01-29}

\title{Gauge invariance in teleparallel gravity theories: \\ A solution to the background structure problem}

\author{E. Minguzzi}
 \affiliation{Dipartimento di Fisica dell'Universit\`a di Milano-Bicocca,\\ Piazza della Scienza 3, 20126 Milano, Italy }

\begin{abstract}
We deal with the problem of identifying a background structure and
its perturbation in tetrad theories of gravity. Starting from a
peculiar trivial principal bundle we define a metric which depends
only on the gauge connection. We find the allowed four-dimensional
structure groups; two of them  turn out to be  the translation
group $T_4$ and the unitary group $U(2)$. When the curvature
vanishes the metric reduces to its background form which coincides
with Minkowski flat metric for the $T_{4}$ case and with the
Einstein static universe metric for the $U(2)$ case. The
perturbation has a coordinate independent definition and allows
for the introduction of observables distinguished from those
obtained from the metric alone. Finally, we show that any
teleparallel theory of gravity, and hence general relativity, can
be considered as a gauge theory over the groups introduced.
\end{abstract}

\maketitle

\section{Introduction}
Teleparallel theories of gravity have a venerable history. In 1928
Einstein \cite{einstein28} introduced  the notion of absolute
parallelism in his attempt of unifying gravity and
electromagnetism. That project failed but later \cite{moller61,
pellegrini62} the idea of a teleparallel geometry was revived as a
geometrical alternative to the Riemmanian approach of general
relativity. Those investigations  led to valuable results
particularly in connection with the study of energy-momentum, its
covariance, positivity and localization \cite{nester89,
deandrade00b} . In 1967 Hayashi and Nakano \cite{hayashi67} showed
that teleparallel gravity can be seen as a gauge theory over the
translation group \cite{cho76, deandrade97a} with a  Lagrangian
quadratic in the strength tensor. Some author started to consider
modifications of the teleparallel equivalent of general relativity
thus discovering a one parameter family of quadratic Lagrangians
which reproduce the correct low energy behavior \cite{hayashi79}.
Such teleparallel theories are experimentally indistinguishable
from Einstein's gravity even if, from the theoretical side, they
are not invariant under local Lorentz trasformations of the tetrad
field. Moreover, the one parameter family cannot be altered with a
slight modification of the coefficients without introducing ghosts
\cite{muller-hoissen85}, and the parameter, for the same reason,
should be taken to be positive. Thus, Kopczy\'nski's
\cite{kopczynski82} proof that such Lagrangians were unable to
determine uniquely the evolution of the teleparallel geometry was
taken as a serious drawback even if, as shown later by Nester
\cite{nester88}, it was less severe than expected and limited to
special solutions. Who takes this problem seriously has to drop
the teleparallel geometry and has to consider it as auxiliary: an
interpretation useful to write down Lagrangians but not realized
in nature. In this case the matter Lagrangian should be invariant
under the hidden symmetries of the gravitational Lagrangian. In
particular the coupling of spinor matter fields with gravity
should be written through the Levi-Civita connection, not
Weitzenb\"ock's \cite{kopczynski82, hayashi81, deandrade01}. This
solution is not completely satisfying because even if we can
introduce a stress-energy tensor of gravity which is covariant
under coordinate transformations, global Lorentz transformations
and gauge transformations \cite{deandrade00b}, we cannot
establish, for instance, the value of the energy density as seen
in a local Lorentz frame. This is because  the stress-energy
tensors so far proposed are not invariant under local Lorentz
transformations of the OT frame, and moreover the dynamics is not
able to define  a privileged OT frame because of the local hidden
Lorentz symmetry of the family of viable Lagrangians
\cite{kopczynski82, hayashi81}. Fortunately, this problem arises
only in those special solutions studied by Nester.
 Alternatively, one can  remove altogether the lack of determinism adding to the
Lagrangian higher order terms in the torsion
\cite{muller-hoissen85}. This, however, not only means a more
radical departure from general relativity (at least from the
theoretical side), but also from the Yang-Mills theory that the
gauge interpretation of teleparallel theory so strongly resembles.

Here we study the teleparallel theories as Yang-Mills theories
even if with a dependence of the Lagrangian on the curvature
somewhat more complicated. Our task is to show that the gauge
formulation of teleparallel gravity theories allows a solution of
the background structure problem. Ultimately this can be
considered as the true reason for the success of these theories in
dealing with the stress-energy tensor. We shall deal only with the
one parameter family of experimentally viable teleparallel
theories, relying on the fact that an unpredictable behavior, if
present, can be removed with a slight perturbation of the initial
value data \cite{nester88}. One can also get rid of it by assuming
that a higher order term in the curvature was implicitly added to
the Lagrangian.

We use the Greek alphabet ($\mu$, $\nu$, $\rho$,\ldots=0,1,2,3) to
denote indices related to spacetime, and the Latin alphabet (a, b,
c,\ldots=0,1,2,3) to denote indices related to the internal space.
The flat spacetime metric $\eta_{a b}$ is fixed by $\eta_{0 0}=1$
and the totally antisymmetric tensor is normalized by $\epsilon_{0
1 2 3}=1$. We use the natural units: $\hbar=1$, $c=1$.

\section{background identification}

The identification of a background structure in theories invariant
under diffeomorphisms is notoriously difficult. This is an
important issue because in turn, in quantum field theory, the
excitations with respect to a fixed background receive a particle
interpretation \cite{maluf94}. Hence the problem of identifying a
background structure is central in quantum gravity. One can also
avoid this problem taking unaltered the spirit of general
relativity. A program of quantization of this kind can be
performed and leads to canonical gravity \cite{ashtekar91}. Before
we abandon the idea of a privileged background structure let us
look more closely to the problem of its definition.

In metric theories of gravity one can try to introduce by hand a
background structure rewriting the metric in the form
\begin{equation} \label{split1}
g_{\mu \nu}=\eta_{\mu \nu} +h_{\mu \nu},
\end{equation}
and identifying $\eta_{\mu \nu}$ with the background  and $h_{\mu
\nu}$ with the perturbation. This can be justified if we are
studying a spacetime which is Minkowskian at spatial infinity but
is still not satisfying. The splitting of the metric in background
and perturbation parts turns out to be dependent on the coordinate
system chosen.  Let two observers label events in different ways
with systems of coordinates $\{x^{\mu} \}$ and $\{ x'{}^{\mu} \}$
respectively. Even if the two systems of coordinates coincide at
spatial infinity, the associated backgrounds $\eta_{\mu \nu} \dd
x^{\mu} \dd x^{\nu}$ and $\eta_{\mu \nu} \dd x'{}^{\mu} \dd
x'{}^{\nu}$ differ.  The prescription (\ref{split1}) introduces a
background in a non-covariant way with respect to coordinate
changes. It cannot be used in many circumstances: for instance in
quantum gravity where we cannot arbitrarily privilege a system of
coordinates \footnote{To privilege a coordinate system  means to
privilege a way  of labeling events. For instance, an observer can
take minus the redshift as  time variable for the light emission
from distant stars, whereas another observer can take, as time
variable, the distance obtained with the distance-redshift
relation. }. We can reach the same conclusion even if we look for
scalars that can be constructed from $\eta$ and $h$. It is not
difficult to show that the only scalars that can be constructed
this way are exactly those that can be constructed using the full
metric $g_{\mu \nu}$ alone. This means, because any observable
should be coordinate independent, that the splitting proposed is
artificial. Any observable expresses properties of the full
geometry, nothing can be said about the background metric and its
perturbation separately. A better definition would be
\begin{eqnarray}
g_{\mu \nu}&=&\bar{g}_{\mu \nu} +h_{\mu \nu}, \\
\bar{R}^{\alpha}_{\ \beta \mu \nu}&=&0.
\end{eqnarray}
It says that a coordinate system  where (\ref{split1}) holds
exists without, however, fixing it from the beginning. The
dynamics should determine that coordinate system  starting from
the initial value data. Unfortunately, this prescription is not
well suited for practical calculations.

 Let us consider a principal fiber bundle $P$ with structure group $G$ and let   the spacetime
$M$ be the base. Over $P$ let us consider a flat connection
$\tilde{\omega}$ and a second connection $\omega$. Then, clearly,
if $M$ is simply connected, the fiber bundle is trivial. A
structure like this is familiar in teleparallel theories where $P$
is the bundle of linear frames and the two connections are given
by Weitzenb\"ock's and Levi-Civita's respectively. Here, however,
this structure acquires a different role. We require $G$ to be a
4-dimensional Lie group and we define the metric of $M$ with
\begin{equation}
g_{\mu \nu}\dd x^{\mu} \dd x^{\nu}=a^{2}I(\tilde{A}-A,
\tilde{A}-A).
\end{equation}
$I$ is a symmetric, bilinear and ad-invariant function defined on
the Lie algebra of the structure group, $I:\mathcal{G} \! \times
\! \mathcal{G} \rightarrow \mathbb{R}$, and $\tilde{A}_{\mu}$,
$A_{\mu}$, are the potentials of the connections in a given
section $\sigma(x)$, e.g.: $A=\tau_{a}A^{a}_{\mu} \dd
x^{\mu}=\sigma^{*}\omega$. The constant $a$ has the dimension of a
length; its presence gives to $A^{a}_{\mu}$ the dimension of a
mass. Under gauge transformations the difference of two potential
transforms with the adjoint representation. Since $I$ is
ad-invariant the metric
so defined is gauge invariant.\\
The function $I$ is a metric for the Lie algebra $\mathcal{G}$. It
must be Minkowkian because $g_{\mu \nu}$ is Minkowskian too. We
can chose  a base of generators $\tau_{a}$ such that
\begin{equation} \label{Inorm}
I(\tau_{a}, \tau_{b})=\eta_{a b}.
\end{equation}
The requirement of ad-invariance for $I$ leads us to the
classification of the allowed four-dimensional groups
\begin{equation}
I([\tau_{a}, \tau_{b}],
\tau_{c})+I(\tau_{b},[\tau_{a},\tau_{c}])=0 \quad \Rightarrow
\quad f_{a b c}=f_{[a b c]},
\end{equation}
where $[\tau_{a},\tau_{b}]=f^{c}_{a b} \tau_{c}$ and $f_{a b
c}=\eta_{a d}f^{d}_{b c}$. Let us rewrite $f_{a b
c}=v^{e}\epsilon_{e a b c}$. With a Lorentz transformation of the
generators we can recast $v^a$ in a canonical form
\cite{weinberg95}, moreover the residual parameter can be
eliminated with a suitable rescaling of $\tau_{a}$, $I$, $a^2$, in
such a way that the product $a^2I$ and Eq. (\ref{Inorm}) are left
unchanged. Finally, the Lie algebras involved are
\begin{itemize}
\item[a)] $v^{a}=0$: Lie algebra of the translation group $T_4$
\begin{equation}
f^{c}_{ a b}=0.
\end{equation}
\item[b)] $v^2>0$: Lie algebra of the group $U(2)$
\begin{equation}
f^{c}_{ a b}=\, 2 \epsilon_{0 a b d} \, \eta^{d c}.
\end{equation}
\item[c)] $v^2<0$: Lie algebra of the group $GL(2, \mathbb{R})$
\begin{equation}
f^{c}_{ a b}= \, 2 \epsilon_{3 a b d} \, \eta^{d c}.
\end{equation}
\item[d)] $v^2=0$: Lie algebra of the two-dimensional Euclidean group with central charge
\begin{equation}
f^{c}_{ a b}=\, \epsilon_{0 a b d} \, \eta^{d c}+\epsilon_{3 a b
d} \, \eta^{d c}.
\end{equation}
\end{itemize}
Notice that the generator $v^{a}\tau_{a}$ belongs to the center of
the group. In the last case it plays the role of  central charge
for the Lie algebra of the Euclidean group in two dimensions
$E_2$.

From Eq. (\ref{Inorm}) it follows that the adjoint representation
acts as a subgroup $H$ of the Lorentz group. Moreover if $v^a \ne
0$, $H$ is the little group of the four-vector $v^{a}$, since the
structure coefficients are left unchanged after an adjoint
transformation. For $G=T_4$ this subgroup is the trivial group
$H=e$, for $G=U(2)$ it is $SO(3)$, for $G=GL(2, \mathbb{R})$ it is
$SO(2,1)$ and in the last case it is $E_2$.

Let $\tilde{\sigma}(x)$ be an horizontal section with respect to
the flat connection $\tilde{\omega}$ and let $\phi(x)$ be the
transition function between the two sections $\sigma(x)
\phi^{-1}(x)=\tilde{\sigma}(x)$ then
$\tilde{A}=\sigma^{*}\tilde{\omega}=\phi^{-1}(x) \dd \phi(x)$.
Finally,
\begin{equation} \label{metric}
g_{\mu \nu}\dd x^{\mu} \dd x^{\nu}=a^{2}I(\phi^{-1} \dd
\phi(x)-A,\, \phi^{-1} \dd \phi(x)-A)
\end{equation}
and the tetrad field can be defined to be
\begin{equation} \label{tetrad}
\tau_{a} e^{a}_{\mu}\dd x^{\mu}=a(\tau_{a}A^{a}_{\mu}\dd
x^{\mu}-\phi^{-1} \dd \phi(x)).
\end{equation}
Notice that $\phi(x)$ is determined only up to global left
multiplications $\phi(x) \rightarrow u \phi(x)$ because of the
arbitrariness of $\tilde{\sigma}(x)$. This does not effect
(\ref{tetrad}) which depends only on the combination  $\phi^{-1}
\dd \phi$.

The degrees of freedom given by the field $\phi$ are easily
removed with a gauge transformation that sends $\phi$ to the
identity. Let us call such gauge "OT gauge" because of its
connection, that we shall exploit, with the OT frame. In the OT
gauge the tetrad field and the potential $A^{a}_{\mu}$ are
proportional. The potential in the OT gauge is our dynamical
variable; in what follows we shall assume that the dynamics
determines this field completely. Clearly, this is not the case if
the Lagrangian is constructed from the metric alone, indeed the
tetrad field is determined only up to a local Lorentz
transformation. Like in teleparallel theories, we have to consider
modifications to the general relativistic Lagrangian.

Let us return to the main question to show that (\ref{metric})
splits the metric in background and perturbation parts in a
coordinate and gauge independent way. We identify the perturbation
with the potential $A$ and the background metric with
\begin{equation}
g_{B} =a^{2}I(\theta,\, \theta),
\end{equation}
where $\theta$ is the canonical 1-form of the group $G$
\cite{kobayashi63}. When the perturbation is a pure gauge
$A=U^{-1} \dd U(x)$ the potential can be sent to zero through a
gauge transformation. Then the metric reduces to $\phi(x)^{*}
g_{B}$ that is, apart from a pullback which amounts simply to a
coordinate transformation \footnote{The coordinate transformation
can be singular. In that case the metric turns out to be
degenerate. This does not happen if the condition $F \approx 0$
 in some region of spacetime follows from the dynamics and if the dynamics itself guarantees the metric to be not degenerate.}    , the metric takes its background form. In other words,
if $F$ is the curvature tensor,
\begin{equation}
F=0 \,\,\Rightarrow \,\,g= \phi^{*} g_{B}.
\end{equation}
The coordinate system in which $g= g_{B}$ is not fixed by the
requirement $F=0$, indeed it depends on the map $\phi(x)$ that in
turn depends on the value of $U(x)$. This shows that the equation
$F=0$ is a coordinate independent way to state the equivalence of
the metric with its background form. Here the splitting is not
artificial because we can construct quantities like $F^{a}_{\mu
\nu}F^{b}_{\alpha \beta} \eta_{a b} g^{\mu \alpha} g^{\nu \beta}$
that are scalars and gauge invariant. Moreover they cannot be
recovered from the metric alone. Their gauge invariance is assured
because the tetrad field (\ref{tetrad}) and the curvature
transform, under gauge transformations, with the adjoint
representation which, as we have seen, is a subgroup of the
Lorentz group.

Before we start studying the dynamics let us esplicitate the
background metric for the structure groups $T_4$ and $U(2)$.
\begin{itemize}
\item[a)]  $T_4$: With the parameterization
$\phi=\eu^{-\tau_{a}\phi^{a}}$ the canonical 1-form becomes
$\theta=\phi^{-1} \dd \phi=-\tau_{a}\dd \phi^{a}$ and the
background metric
\begin{equation*}
g_{B}=a^{2}\eta_{a b} \,\dd \phi^{a} \dd \phi^{b}
\end{equation*}
coincides with the Minkowski metric.
\item[b)] $U(2)$: With the parameterization
\begin{equation*} \label{param}
\phi=\eu^{\lambda \tau_{0}} \,\eu^{\chi(\tau_{1} \sin \theta \cos
\varphi +\tau_{2} \sin \theta \sin \varphi +\tau_{3} \cos \theta)}
\ ,
\end{equation*}
and the representation (the metric is independent from the
representation of the Lie algebra chosen) $\tau_{\mu}=i
\sigma_{\mu}$, $\sigma_{0}=\mathrm{I}$, we find $I(\alpha,
\alpha)=-\mathrm{det}(\alpha)$, $\alpha \in \mathcal{G}$ and
\begin{equation*}
\,\, g_{B}\!=\! a^{2} \{ \dd \lambda^{2}-  \dd
\chi^{2}-\sin^{2}\chi (\dd \theta^{2}+\sin^{2} \theta \, \dd
\varphi^{2})  \}.
\end{equation*}
Hence, for $G=U(2)$ the background metric coincides with that of
Einstein's static universe.
\end{itemize}
Analogous calculations lead to the background metric for the cases
c) and d). They, however, do not enjoy the cosmological principle.

To summarize, we have identified the perturbation with the
potential of a gauge theory. Its transformation under gauge and
coordinates change are well known. From that variable one can
construct scalars that express properties of the perturbation with
respect to a background structure. The scalars to be considered
depend on the supposed background structure, for instance, they
are invariant under $U(2)$ gauge transformations in an Einstein's
static universe background.

\section{Dynamics and teleparallel theories}
Now we have to construct a dynamics for the gauge potential.
 Let $\hat{A}^{a}_{\mu}$ be the
potential in the OT gauge and let $\hat{e}^{a}_{\mu}$ be   the
tetrad field (\ref{tetrad}) in the same gauge:
$\hat{e}^{a}_{\mu}=a \hat{A}^{a}_{\mu}$. We introduce a
teleparallel geometry on $M$ of OT frame $\{\hat{e}^{a}_{\mu}\}$,
where  the brackets  recall that the OT frame is defined up to a
global Lorentz transformation. We can rewrite  the usual
expression of the torsion in terms of the OT frame
\begin{equation}
T^{\rho}_{\mu \nu}=\hat{e}^{\rho}_{a}(\p_{\mu} \hat{e}^{a}_{\nu}
-\p_{\nu}\hat{e}_{\mu}^{a}),
\end{equation}
in a gauge invariant way
\begin{equation} \label{torsion}
T^{\rho}_{\mu \nu}=a \,e^{\rho}_{c} F^{c}_{\mu \nu}-\frac{1}{a}\,
e^{\rho}_{c} f^{c}_{a b} \,e^{a}_{\mu} \, e^{b}_{\nu},
\end{equation}
where the tetrad field is given by Eq. (\ref{tetrad}) and where
\begin{equation}
F^{a}_{\mu \nu}=\p_{\mu} A^{a}_{\nu} -\p_{\nu}A_{\mu}^{a}+f^{a}_{
b c} \,A^{b}_{\mu} A^{c}_{\nu}.
\end{equation}
In order to shorten the notation let $F_{a b c}=\eta_{a d}
F^{d}_{\mu \nu} e^{\mu}_{b} e^{\nu}_{c}$ and $T_{a b c}=e_{c \rho}
T^{\rho}_{\mu \nu} e^{\mu}_{b} e^{\nu}_{c}$. Eq. (\ref{torsion})
becomes
\begin{equation}
T_{a b c}= a F_{a b c}-\frac{1}{a} \,f_{a b c},
\end{equation}
and the one parameter family of viable teleparallel Lagrangians
\cite{hayashi79, muller-hoissen85}
\begin{eqnarray} \label{tellag}
\mathcal{L}&=&\frac{\sqrt{-g}}{16 \pi G} \,\,\{\frac{1}{4} T_{\alpha \mu \nu}
T^{\alpha \mu \nu} +\frac{1}{2} T_{\alpha \mu \nu} T^{\mu \alpha
\nu}+ \nonumber \\
&-&T^{\mu}_{\ \mu \alpha} T^{\nu \ \alpha}_{\ \nu} + \alpha
\frac{3}{2}T_{[\alpha \mu \nu]}T^{[\alpha \mu \nu]} \},
\end{eqnarray}
 becomes
\begin{eqnarray} \label{gaugelag}
\mathcal{L}&=&\frac{ \sqrt{-g}}{16 \pi \sigma^2}\,
\left\{\frac{1+2\alpha}{4} F_{a b c} F^{a b c}+\frac{1-2 \alpha}{2}F_{a b c} F^{b a c}+\right. \nonumber \\
&{}&\!\!\!\!\!\!\!\!\!\!\!\!\!\!\!\!-\left.F^{a}_{\ a b} F^{c \
b}_{\ c}+ \frac{1-6 \alpha}{2} F_{a b c} f^{a b c} -\frac{1-6 \alpha}{4 a^{2}}f_{a b c} f^{a b c} \right\}.
\end{eqnarray}
The linearized theory is ghost-free when the parameter $\alpha$ is
positive \cite{hayashi79}: $\alpha \ge 0$. General relativity
corresponds to $\alpha=0$. The first term in Eq. (\ref{gaugelag})
is the well known Yang-Mills Lagrangian; its dimensionless
coupling constant is given by
\begin{equation} \label{coupling}
\sigma=\frac{L_{P}}{a}.
\end{equation}
The reader should keep in mind that this similarity is only
apparent: here even the metric and the tetrad field depend on the
potential A; this leaves us with a theory that is still not
renormalizable. The role of Eq. (\ref{gaugelag}) is to exhibit the
gauge invariance of teleparallel gravity theories, and hence of
general relativity,  under any of the four dimensional group
studied in the previous section. Actually, we  have obtained this
gauge invariance with the introduction of a new degree of freedom
given by the field $\phi(x)$. One can correctly suspect that the
need for the introduction of a new field is a signal  that the
symmetries developed are not genuine new symmetries of the
original Lagrangian. Moreover, one can be concerned about the fact
that our dynamical variable is the potential in the OT gauge: if
in practical calculation we have to return to the usual tetrad
formulation, what is the advantage of introducing such dependence
of the metric? To answer these questions we need to introduce the
internal coordinate representation.

So far, we have used the spacetime coordinates. Let us choose  a
gauge such that $\phi: M \rightarrow G$ is injective. In such a
gauge we can perform a coordinate transformation from the
spacetime coordinates $\{x^{\mu}\}$ to  given internal coordinates
$\{ \phi^{a} \}$ of the group manifold $G$. For example the
Lagrangian in the internal coordinates becomes $\phi^{-1
*}\mathcal{L}(x)$ with a tetrad field  given by \footnote{For
clearness, in internal coordinates the reciprocal of $e^{a}_{b}$
should be written ${e^{-1}}^{b}_{c} \,$. }
\begin{equation} \label{tetrad2}
\tau_{a} e^{a}_{b}(\phi^{a}) \,\dd \phi^{a}=\tau_{a}
A^{a}_{b}(\phi^{a}) \, \dd \phi^{b}-\theta.
\end{equation}
From this equation we see that the field $\phi(x)$ has been
completely removed by the coordinate change. In the internal
coordinate representation the gauge symmetry turns out to be an
alternative way of recasting the invariance under coordinate
transformations. To any coordinate change in the internal
formalism corresponds a gauge transformation in the spacetime
formalism; the converse, however, is not true. For instance the OT
gauge can not be accomplished in the internal formalism because
the field $\phi(x)$, in that case,  is not injective. In the
internal representation the invariance under coordinate
transformation and the invariance under gauge transformation are
linked: the transformation law for the potential becomes
($\sigma'=\sigma\, u$)
\begin{equation} \label{inttrasf}
\tau_{a} {A'}^{a}_{c} =\{u^{-1}\tau_{a} A^{a}_{b}u+u^{-1} \p_{b}
u\} \frac{\p \phi^{b}}{\p {\phi'}^{c}}\, ,
\end{equation}
and the transformation law for the curvature becomes
\begin{equation}
F'_{a b}=u^{-1}F_{c d}\, u \,\frac{\p \phi^{c}}{\p {\phi'}^{a}}
\frac{\p \phi^{d}}{\p {\phi'}^{b}}\, ,
\end{equation}
where the matrix $u(\phi)$ is related to the transformation
${\phi'}^{ a}(\phi^{b})$ by the product $\phi'=\phi \,u(\phi)$. In
the same way it can be shown, for example,  that the metric given
by (\ref{metric}) transform as a tensor under (\ref{inttrasf}).
Notice that in the internal formalism there is no distinction
between internal and spacetime indices because in the spacetime
manifold we have introduced internal coordinates.

Let us investigate more closely the case $G=T_{4}$. In the
previous section we defined the coordinates $\{ \phi^{a} \}$ on
the group manifold: $\phi= \eu^{-\tau_{a} \phi^{a}}$. The tetrad
field in the spacetime formalism is \cite{hayashi67}
\begin{equation}
e^{a}_{\mu}=\p_{\mu}\phi^{a}(x)+A^{a}_{\mu}(x),
\end{equation}
whereas in the internal formalism is (Eq. (\ref{tetrad2}))
\begin{equation}
e^{a}_{b}(\phi)=\delta^{a}_{b}+A^{a}_{b}(\phi).
\end{equation}
Apart for changes due to the notation, this is exactly the
dependence of the tetrad field on the potential given in
\cite{cho76}. Indeed, our formalism in the $G=T_4$ case gives rise
to the widely studied translational gauge symmetry of teleparallel
theories. In literature it has been studied both in  internal
coordinates \cite{cho76} and in spacetime coordinates
\cite{hayashi67, deandrade00b}.

Now we see the advantage of these gauge formulations. In the
internal coordinate representation, contrary to what happens in the spacetime representation, the Lagrangian does not  reduce to its
usual tetrad form because we are not working (and we cannot work)
in the OT gauge. Moreover, the dynamical variable is given by the
potential $A^{a}_{b}(\phi)$ and there is no further degree of
freedom apart from that of making gauge-coordinate
transformations. The metric depends on the potential through the
tetrad field (\ref{tetrad2}). The identification of the potential
$A(\phi)$ with the perturbation is coordinate independent, its
transformation law being Eq. (\ref{inttrasf}).

As a final point let us investigate the invariance under global
Lorentz transformations. We should expect that the formulation in
terms of a perturbation spoils this explicit invariance see (Eq.
(\ref{tellag})) if the background structure does not share the
same symmetry. This is indeed the case. Only the structure
coefficients of the group $G=T_{4}$ are left unchanged under the
replacement $\tau_{a}=\Lambda^{b'}_{\ a} \tau_{b'}$. For this
group one can accomplish the global Lorentz invariance with
$\phi^{a}=\Lambda^{a}_{\ b'} \phi^{b'}$ and $
A^{a}_{\mu}(x)=\Lambda^{a}_{\ b'} A^{b'}_{\mu}(x)$. Of course, in
the internal representation even the coordinates transform and we
have
\begin{equation}
A^{a}_{b}(\phi)=\Lambda^{a}_{\ c'} \, A^{c'}_{d'}(\phi)\,
\Lambda^{d'}_{\ b}.
\end{equation}
Surprisingly, in the previous literature, despite of the
properties of the gauge formulation,  most of the calculations
were performed in the usual tetrad formalism. This, however, is
quite natural if one looks for non-perturbative results.

\section{Conclusions}
In order to solve the background structure problem of gravity
theories, we proposed  a metric of the form (\ref{metric}) where
two connections were introduced.  We showed that, if the dynamics
determines the potential of the gauge theory completely (up to
gauge or coordinate changes), the splitting of the metric in
background and perturbation parts is physical since new
observables, not dependent from the full metric alone, can be
constructed. We classified the allowed four-dimensional structure
groups finding suitable backgrounds for the $T_4$ and $U(2)$
cases. We went to exploit the dynamics with the introduction of a
teleparallel geometry on the manifold. Taking advantage of
teleparallel gravity results, that geometry was used to construct
Lagrangians compatible with the low energy limit of general
relativity. It should be mentioned, however, that the gauge
approach allows a more large variety of gravitational Lagrangians
than the teleparallel approach. For instance the Lagrangian
$F^{a}_{\mu \nu}F^{b}_{\alpha \beta} \delta_{a b} g^{\mu \alpha}
g^{\nu \beta}$ is coordinate and gauge independent but has no
teleparallel analogue. Teleparallel Lagrangians have the advantage
of being invariant under global Lorentz transformation. This still
seems a necessary condition for the introduction of spinor matter
fields. Finally, teleparallel theories were interpreted as gauge
theories over different backgrounds in dependence of the structure
group chosen. Most interesting cases were the Minkowski flat
metric and the Einstein static universe. As a final comment, Eq.
(\ref{coupling}) suggests that renormalization can effect the
'radius' of the universe. Of course, the theory is not
renormalizable and this is  only a speculation, anyway it seems to
deserve further investigations.

\begin{acknowledgments}
I wish to thank C. Destri for useful discussions and for a
critical reading of the manuscript,  and INFN for financial
support.
\end{acknowledgments}

\end{document}